\documentclass[12pt]{article}
\usepackage{cjp}
\newcommand{\beq}{\begin{equation}}
\newcommand{\eeq}{\end{equation}}
\newcommand{\beqn}{\begin{eqnarray}}
\newcommand{\eeqn}{\end{eqnarray}}

\newcommand{\CPV}{\overline{CP}}

 \def\bc{\begin{center}} \def\ec{\end{center}}
\def\be{\begin{equation}}
\def\ee{\end{equation}}
\def\bea{\begin{eqnarray}}
\def\eea{\end{eqnarray}}

\newcommand{\lqcd}{\Lambda_{\rm QCD}}

\def\dirule{$\Delta I=1/2$ rule}
\def\oplus{\widehat{O}^{(+)}}
\def\ominus{\widehat{O}^{(-)}}
\def\oplmi{\widehat{O}^{(\pm)}}

\def\hat{\widehat}
\def\tilde{\widetilde}

\newcommand{\<}{\langle}
\renewcommand{\>}{\rangle}

\begin{document}
\title{Lattice QCD, O.P.E. and the Standard Model}
\author{G.C.~Rossi}
\address{Dipartimento di Fisica, Universit\'{a} di Roma
``Tor Vergata''\\ INFN Sezione di Roma II\\
Via della Ricerca Scientifica 1, 00173 Roma, 
 ITALY\\ rossig@roma2.infn.it}
\maketitle
\begin{abstract}
A number of old and new methods for computing $K\to\pi\pi$ amplitudes on the 
lattice are reevaluated. They all involve a non-perturbative determination
of matching coefficients. I will show how problems related to operator 
mixing can be greatly reduced by introducing the O.P.E. of hadronic currents
directly on the lattice. Applications to the evaluation of CP-symmetric
($\Delta I = 1/2$ rule) and CP-violating ($\epsilon '/\epsilon$) processes
are presented.
\end{abstract}

\section{Introduction}
\label{sec:introd}

The Kaon sector of the Standard Model may still have some unexpected
surprise in store for us. Uncovering them requires a very good
non-perturbative knowledge of hadronic matrix elements. Despite the many
non-perturbative approaches developed to this aim (chiral
lagrangians, $1/N$-expansions,...)~\footnote{Excellent reviews can
be found in refs.~\cite{rev1}--\cite{rev5}.}, it seems
that only lattice based computations can cope with this formidable problem.

There are two major difficulties, however, which arise in the calculation 
of hadronic matrix elements in lattice QCD. 
\begin{enumerate}
\item
Decay amplitudes into two or more particles cannot be directly accessed 
in Euclidean space, as a consequence of the Maiani and Testa (MT) no-go
theorem~\cite{MT}, except at threshold, where, we recall, final state
interactions are absent.   
\item
Operators can mix with operators of lower dimension with coefficients
which diverge as inverse powers of the lattice spacing. These
contributions must be computed non-perturbatively and subtracted,
most often leading to prohibitively large statistical errors~\cite{gavela}. 
\end{enumerate}

In this talk I will be mainly concerned with the second of these two
problems. In Section~\ref{sec:deltai} I recall some of the old methods
proposed to deal with the question of operator mixing in the case of Wilson
fermions (either improved or not)~\footnote{For the use of staggered
fermions see ref.~\cite{toolkit}. For the new proposals based on
Ginsparg-Wilson, overlap or domain wall fermions see the many interesting
contributions to these Proceedings.}. In the discussion I will make
explicit reference to the problem of explaining the large value of the ratio
between the $I=0$ and the $I=2$ $K\rightarrow\pi\pi$ amplitudes ($I$ is the
isospin of the two pions in the final state). In Section~\ref{sec:ope} I will
illustrate a new approach based on the direct use of the O.P.E. on the
lattice~\cite{noinew}. The idea is to measure in Monte Carlo simulations the
small $x$-behaviour ($a\ll |x| \ll \Lambda_{\rm QCD}^{-1}$, $a$ lattice
spacing) of the hadronic matrix elements of the product of two hadronic
currents and to compare it with the $x$-behaviour of its  O.P.E. In the
O.P.E. formula one should look at the Wilson coefficients as known functions
(assumed to have been computed in perturbation theory), while the matrix
elements of the corresponding operators are treated as fitting
parameters. If the condition $a\ll |x| \ll \lqcd^{-1}$ is satisfied,
perturbation theory applies, up to corrections vanishing as some power of
$a/|x|$, and the best-fit values of the above parameters will automatically
supply the physical matrix elements of the finite, renormalized (continuum)
operators without the need of knowing their expression in terms of bare
lattice operators. A feasibility study of this method in the case of the
two-dimensional O(3) $\sigma$-model has been successfully carried out in
ref.~\cite{MONT}.
Applications of the general philosophy underlying this approach to the
lattice computation of CP-violating $K\rightarrow \pi\pi$ amplitudes,
relevant for the evaluation of
$\epsilon'/\epsilon$~\cite{rev1}--\cite{rev5}, will be presented in
Section~\ref{sec:eps}. 
For lack of space I will not discuss here a recent interesting extension of
the above ideas, in which also the Wilson coefficients are extracted in a
non-perturbative way from lattice data~\cite{sch}. In this stronger 
formulation the O.P.E. method has been already concretely applied by the
authors of ref.~\cite{sch} to the computation of the first moment of D.I.S.
structure functions.
Conclusions and an outlook on the perspectives of using the lattice O.P.E.
method and of its potentialities and limitations can be found in
Section~\ref{sec:concs}.

\section{The old approaches to the $\Delta I= 1/2$ rule}
\label{sec:deltai}

One of the major puzzles remaining in hadronic physics is the
origin of the so-called ``$\Delta I=1/2$ rule" in non-leptonic kaon decays.
Decays in which isospin changes by $\Delta I=1/2$ are greatly
enhanced over those with $\Delta I=3/2$. One finds
experimentally (neglecting CP violation)
\begin{equation}
{{\cal A}(K\to\pi\pi[I=0]) \over {\cal A}(K\to\pi\pi[I=2]) } \approx 20 \,.
\label{eq:dirule}
\end{equation}
Although the origin of this large enhancement is not theoretically well
understood, we do know that, in a QCD-based explanation, most of the
enhancement must come from long distance, non-perturbative physics, or else
new physics is winking at us here!

Let me briefly discuss the source of the difficulties in calculating ${\cal
A}(K\to\pi\pi)$ in lattice QCD~\cite{ancient}~\cite{gavela}.

For scales below $M_{_W}$, but above the charm quark mass, the $\Delta S=1$
part of the effective weak Hamiltonian, ${\cal H}_{\rm eff}^{\Delta S=1}$,
can be written in the form  
\bea
{\cal H}_{\rm eff}^{\Delta S=1}
&=&
\lambda_u {G_F \over \sqrt2} 
\left[ C_+(\mu, M_{_W}) \oplus(\mu) + C_-(\mu,M_{_W}) \ominus(\mu) \right]\,,
\label{eq:HW}\\
O^{(\pm)} &=& \left[ (\bar s \gamma_\mu^L d)(\bar u \gamma_\mu^L u)
 \pm (\bar s \gamma_\mu^L u)(\bar u \gamma_\mu^L d) \right]
  -  \left[ u \leftrightarrow c \right] \,,
\label{eq:oplmidef}
\eea
where $\gamma_\mu^L=\gamma_\mu (1\!-\!\gamma_5)/2$, 
$\lambda_u=V_{ud}V_{us}^*$, $G_F$ is the Fermi constant and $\mu$ is the
subtraction point. Throughout this paper I will denote finite
renormalized operators with a $\,\widehat{}\,$ on top.

The operators $\widehat{O}^{(+)}$ and $\widehat{O}^{(-)}$ have different
transformation properties under isospin. In particular, $\ominus$ is pure
$I=1/2$, whereas $\oplus$ contains parts having both $I=1/2$ and $I=3/2$.
An explanation of the \dirule\ thus requires
that the $K\to\pi\pi$ matrix element of $C_- \ominus$ 
be substantially enhanced compared  to that of $C_+ \oplus$.

Part of the enhancement is provided by the ratio of Wilson coefficients,
$C_-/C_+$, in their renormalization group evolution from $\mu\sim M_{_W}$ 
down to $\mu \sim 2\;$GeV, where one finds $|C_-/C_+| \approx 2$.
This factor is, however, too small by an order of magnitude to explain the
\dirule. The remainder of the enhancement must come from the matrix elements
of the operators, and these are the quantities that we wish to evaluate
on the lattice. Attempts in this direction date back to the works of 
ref.~\cite{ancient}

There is a long list of methods proposed in the literature to deal with the
problem of computing  on the lattice the hadronic matrix elements that enter 
in the non-leptonic weak decay amplitude. They are all aimed at bypassing the
two major difficulties mentioned in the Introduction. In this section I wish
to briefly highlight the merits and the drawbacks of the most promising
among them. Before doing that let me start by recalling what is the  mixing
pattern of the operators $O^{(\pm)}$.

\subsection{The mixing of $O^{(\pm)}$ on the lattice}
\label{sec:deltai1}

As is well known, the explicit breaking of chiral symmetry due to the 
presence of the Wilson term in the lattice action induces the mixing of the
bare operators $O^{(\pm)}$ with operators belonging to different chiral
representations. Taking into account the symmetries left unbroken by the
lattice regularization, it can be shown that, in terms of bare operators, the
renormalized, finite lattice operators possessing the correct chiral
transformation properties have the form~\cite{boc}   
\beq
\widehat{O}^{(\pm)}=\widehat{O}_{PC}^{(\pm)}+\widehat{O}_{PV}^{(\pm)}
\label{PCPV}\eeq
with
\beqn
\begin{array}{l}
\widehat{O}_{PC}^{(\pm)}=Z_{PC}^{(\pm)}[O^{(\pm)}_{PC}+\sum_{i=1}^4
C^{(\pm)}_{iPC} O^{(\pm)}_{iPC}+C^{(\pm)}_{\sigma F}
\bar{s}\sigma_{\mu\nu}F_{\mu\nu}d + C^{(\pm)}_{\bar{s}d} \bar{s}d]
\\ \\
\widehat{O}_{PV}^{(\pm)}=Z_{PV}^{(\pm)}[O^{(\pm)}_{PV}+C^{(\pm)}_{\sigma
\widetilde{F}} \bar{s}\sigma_{\mu\nu}\widetilde{F}_{\mu\nu}d +
C^{(\pm)}_{\bar{s}\gamma_5 d} \bar{s}\gamma_5 d]\, .
\end{array}\label{eq:RINORM} \eeqn
$O^{(\pm)}_{PC}$ and $O^{(\pm)}_{PV}$ are respectively the
parity-conserving (PC) and the parity-violating (PV) parts of the operators
$O^{(\pm)}$. The $O^{(\pm)}_{iPC}$'s (i=1,...,4) are (6 dimensional)
four-quark operators whose explicit expression can be found
in~\cite{DGMSTV}. As we have indicated above, parity-conserving and
parity-violating operators renormalize separately, because strong
interactions conserve parity. 

The mixing pattern shown in eqs.~(\ref{eq:RINORM}) is the same both with the
standard Wilson action and with the ``tree-level" clover-improved 
action~\cite{SWH}, provided in the latter case quark fields are 
appropriately improved (``rotated"), according to
$q \rightarrow (1+a m_0/2) q$, $\bar{q}\rightarrow (1+a m_0/2) \bar{q}$,
with $m_0$ the bare quark mass parameter entering in the fermion action.
If the action is improved non-perturbatively, i.e. beyond 
tree-level~\cite{luscher}, the form of eqs.~(\ref{eq:RINORM}) will
become much more complicated because of the presence of many more operators, 
needed to correct all higher orders in $a$, beyond the leading $a
g_0^{2n}\log^n a$ terms already taken care of by the tree-level Pauli term
in the clover action. To my knowledge a complete analysis of the mixing in
this case is still lacking.

Let us now examine in turn the various terms in eqs.~(\ref{eq:RINORM}).

$\bullet$ {\underline{Dimension 6 operators}}

Spin and color structure of the operators of dimension 6 contributing
here is the same as that of the $\Delta S=2$ case discussed
in ref.~\cite{DGMSTV}. Only the flavor structure is different. Notice that
there is no mixing with dimension 6 operators in the parity-violating part. 

$\bullet$ {\underline{Dimension 5 operators}}

Thanks to the GIM cancellation mechanism, the coefficients
$C^{(\pm)}_{\sigma F}$ and $C^{(\pm)}_{\sigma \widetilde{F}}$ are actually
finite, because the potential $1/a$ divergence in the matrix elements 
of $O^{(\pm)}_{PC}$  and $O^{(\pm)}_{PV}$ gets replaced by a
$m_c-m_u$ factor. Furthermore the CPS symmetry~\cite{direct} (CPS = CP
$\times$ symmetry under $s\rightarrow d$ exchange) makes $C^{(\pm)}_{\sigma
\widetilde{F}}$ proportional to $m_s-m_d$ and thus vanishing in the limit of
exact vector flavor symmetry (exact $SU(N_f)_V$).

$\bullet$ {\underline{Dimension 3 operators}} 

The GIM mechanism softens the a priori possible $1/a^3$ divergence of
$C^{(\pm)}_{\bar{s}d}$ and $C^{(\pm)}_{\bar{s}\gamma_5 d}$, because, as I
said, it has the effect of substituting one factor $1/a$ with
$m_c-m_u$. In the case of $C^{(\pm)}_{\bar{s}\gamma_5 d}$ an extra $1/a$
power is taken care of by CPS symmetry which has the effect of
replacing it by a $m_s-m_d$ factor.

\subsection{CP-conserving $K\rightarrow\pi\pi$ amplitudes}
\label{sec:deltai2}
In this section I wish to briefly describe a number of methods that have
been proposed to deal with the problem of computing CP-conserving 
$K\rightarrow\pi\pi$ amplitudes on the lattice.

1. In ref.~\cite{direct} it was suggested to work with $m_s=m_d$ and
calculate the amplitudes  
\be 
{\cal A}^{(\pm)} = \langle \pi(\vec p_1\!=\!0)
\pi(\vec p_2\!=\!0)|\widehat{O}^{(\pm)}(\mu)|K(\vec p_K\!=\!0)
\rangle\bigg|_{m_s=m_d}  
\label{eq:b+s} \ee
with all three particles at rest. We have seen that, setting $m_s=m_d$, GIM
cancellation and CPS-symmetry cause all mixings in the parity-violating
part to vanish, removing the need for subtractions altogether. Since one is
working at threshold with the two pions at rest, the MT no-go 
theorem~\cite{MT} does
not apply. So both difficulties mentioned in the Introduction are overcome
with this simple choice. The method requires, however, a large extrapolation
from the unphysical, off-shell point, $m_s=m_d$, to the physical one with the
use of chiral perturbation theory~\cite{GL}. This procedure can be
particularly delicate if lattice numbers are extracted from quenched
simulations, due to the presence of chiral logarithms~\cite{CHLOG}.

2. An alternative method~\cite{noinew} consists in working with the
non-perturbatively $O(a)$ improved clover action (for which there are no
errors of $O(a)$ in the spectrum~\cite{SWH} and on-shell matrix elements of
improved currents obey the continuum chiral Ward identities up to
$O(a^2)$~\cite{luscher}). Choosing quark masses such that  $m_K=2m_\pi$, one
measures the $K\to\pi\pi$ amplitudes again with all particles at rest. Since
in these kinematical conditions, the total momentum transfer vanishes
($\Delta p=0$), one can prove that the matrix element of the dangerous,
power divergent, subtraction of the operator $\bar{s}\gamma_5 d$ is now of
$O(a)$ rather than $O(1)$ and vanishes in the continuum limit. Furthermore 
pions are at rest and again the MT no-go theorem does not apply. Besides the
need of a (perhaps less severe) chiral extrapolation (unlike the previous
case here we are dealing with on-shell amplitudes), a problem with this
method is the difficulty of tuning quark masses with a sufficiently high
accuracy to have the condition $m_K= 2m_\pi$  satisfied. This problem can
be, however, alleviated by explicitly performing the subtraction of the
$\bar s \gamma_5 d$ operator. 

3. The old proposal of ref.~\cite{MMRT} is based on the evaluation of the
$K\to\pi$ matrix elements of (the positive parity part of) the weak
Hamiltonian and makes use of chiral perturbation theory, in the form of Soft
Pion Theorems (SPT's), to connect $K\to\pi$ to $K\to\pi\pi$ amplitudes.
Since only single-particle states are involved, there are no problems with
the MT no-go theorem. The disadvantage of the method is that, as 
discussed before, the operator mixing problem for the positive parity part of
$O^{(\pm)}$ is much more severe than for their negative parity part. This
makes an accurate evaluation of the matrix elements of the renormalized
operators very difficult.

The relation between $K\to\pi\pi$ and $K\to\pi$ amplitudes can be found
from the classical works of refs.~\cite{SPTS}. At leading order in chiral
perturbation theory the physical amplitude takes the form (for $\Delta p=0$) 
\be
\<\pi^+\pi^-\vert  \widehat{O}^{(\pm)}(\mu)\vert K^0\> =
i\,\gamma^{(\pm)}\, {m^2_K - m^2_\pi\over f_\pi} 
\,. \label{eq:SPT3}
\ee
As the coefficients $\gamma^{(\pm)}$ appear also in the expression for
the $K\to\pi$ matrix element 
\begin{equation}
\<\pi^+(p)\vert  \widehat{O}^{(\pm)}(\mu) \vert K^+(q)\> =
-\delta^{(\pm)}\, {m^2_K\over f_\pi^2} + \gamma^{(\pm)}\,p\cdot q
\, ,\label{eq:SPT2}
\end{equation}
by studying this matrix element on the lattice as a function of $p\cdot q$, 
one can, in principle, determine $\gamma^{(\pm)}$, from which one then
obtains the $K\to\pi\pi$ amplitudes~\footnote{For completeness I recall
that, in the same notations used in eqs.~(\ref{eq:SPT3}) and~(\ref{eq:SPT2}),
one has $\<0\vert\widehat{O}^{(\pm)}\vert K^0\>
=i(m_K^2-m_\pi^2)\delta^{(\pm)}$.}.

In order to construct the finite renormalized lattice operators 
$\widehat{O}^{(\pm)}$ it was suggested in reference~\cite{MMRT} to use
perturbation theory to determine the finite mixing coefficients
($C^{(\pm)}_{iPC}$ and $C^{(\pm)}_{\sigma F}$) and subtract 
non-perturbatively the operator $\bar s d$, as its mixing coefficient,
$C^{(\pm)}_{\bar{s}d}$ (see the first of eqs.~(\ref{eq:RINORM})), is
quadratically divergent. This approach has been tried in ref.~\cite{gavela}
with no success as numerical and statistical errors coming from the power
divergent subtraction completely obscure the physical signal.

\section{Lattice O.P.E.}
\label{sec:ope} 

In this section I want to illustrate a strategy which, in principle,
avoids all the difficulties caused by mixing with lower dimension operators
and automatically gives the physical matrix elements of the effective weak
Hamiltonian with the correct normalization~\cite{noinew}. 

The method is based on the idea of studying at short distances ($a\ll |x| 
\ll  \lqcd^{-1}$) the $x$-behaviour of the O.P.E. of two
hadronic currents on the lattice. It does not use chiral perturbation
theory, and thus in principle applies equally well to the $\Delta S=1$,
$\Delta C=1$ and $\Delta B=1$ parts of the weak Hamiltonian. In addition, it
allows one to construct an improved weak Hamiltonian (i.e.~one having errors
of $O(a^2)$), if the improved version of the weak hadronic
currents~\cite{luscher} is used.
The approach is speculative in the sense that it is likely to require more
computational power than is presently available, although it may
become practical with the advent of Teraflop machines.

I recall that the standard construction of the non-leptonic weak Hamiltonian
begins with the formula 
\begin{equation}
{\cal H}_{\rm eff}^{W}=\frac{g^2_{_W}}{8}\int d^4x\,D_{_W} (x;M_{_W}) 
T\Big(J_{\rho L}(x) J^\dagger_{\rho L}(0)\Big) \, ,
\label{eq:HEFF}
\end{equation}
where
\begin{equation}
D_{_W}(x;M_{_W})=\int\frac{d^4p}{(2\pi)^4}
\,\frac{\mbox{e}^{ipx}}{p^2+M_{_W}^2}  
\label{WPROP} 
\end{equation}
is the longitudinal part of the $W$-boson propagator and $J_{\rho L}$ is the
(left-handed) hadronic weak current. One then introduces the Wilson operator
product expansion in the r.h.s of eq.~(\ref{eq:HEFF}), a step which is
justified by the observation that the dominant contribution to the integral
comes from very small distances, $|x| \ll M_{_W}^{-1}$. For physical
amplitudes, one obtains in this way 
\begin{equation}
\langle h|{\cal H}_{\rm eff}^{W}|h'\rangle =
\frac{G_{F}}{\sqrt{2}} \sum_i C_i(\mu,M_{_W}) M_{_W}^{6-d_i} 
\langle h|\widehat{O}^{(i)}(\mu)|h'\rangle\ ,
\label{eq:HEFFOPE}
\end{equation}
where $|h \rangle$ $|h'\rangle$ are hadronic states, $d_i$ is the dimension
of the operator $\widehat{O}^{(i)}(\mu)$ and
$G_{F}/\sqrt{2}=g^2_{_W}/{8M_{_W}^2}$. The functions $C_i(\mu,M_{_W})$
result from the integration of the Wilson expansion coefficients,
$c_i(x;\mu)$ (defined in  eq.~(\ref{eq:ME}) below), with the $W$-propagator
\begin{equation} C_i(\mu,M_{_W}) M_{_W}^{6-d_i} 
= \int d^4x\,D_{_W} (x;M_{_W}) c_i(x;\mu)\ .
\label{eq:WILCOEF}
\end{equation}
The $\widehat{O}^{(i)}(\mu)$'s are quark and/or gluon operators renormalized
at the subtraction point $\mu$. The functions $C_i(\mu,M_{_W})$ are evaluated
in perturbation theory and their running with $\mu$ is dictated by the
renormalization group equation which follows from the $\mu$-independence of
the l.h.s. of eq.~(\ref{eq:HEFFOPE}). The sum in the
expansion~(\ref{eq:HEFFOPE}) is over operators of increasing dimension. As
the operator dimension of $\cal {H}_{\rm eff}^{W}$  is 6, we will have to
consider in the following only operators with dimensions  $d_i \le 6$, since
the contribution from operators with $d_i>6$ is suppressed  by powers of
$1/M_{_W}$.

All the intricacies and complications of operator mixing in the definition of
the finite and renormalized operators, $\widehat{O}^{(i)}(\mu)$, come about
because the integrals in~(\ref{eq:HEFF}) and~(\ref{eq:WILCOEF}) are extended
down to the region of extremely small $|x|$. The complicated mixing pattern
of the $\widehat{O}^{(i)}(\mu)$'s in terms of bare operators arises from
contact terms when the separation of the two currents goes to zero,
i.e.~when $|x|$ is of the order of $a$. This observation suggests that a
simple way to circumvent these difficulties is to directly determine the
matrix elements of renormalized operators by enforcing the validity of
the O.P.E. on the lattice for distances $|x|$ much larger than the lattice
spacing $a$, but much smaller than $\lqcd^{-1}$ (and inverse quark masses),
i.e.~in a region where perturbation theory determines the form of the Wilson
expansion. 

We should imagine proceeding in the following way. If $J_{\rho L}$ is the 
appropriately renormalized (and possibly improved) finite lattice current
operator, one starts by measuring in a Monte Carlo simulation the hadronic
matrix element  $\<h\vert T(J_{\rho L}(x) J^\dagger_{\rho L}(0))\vert h'\>$,
as a function  of $x$ in the region  $a\ll |x|  \ll  \lqcd^{-1}$. The
numbers $\< h\vert \widehat{O}^{(i)}(\mu)\vert h'\>$
(eq.~(\ref{eq:HEFFOPE}))  are extracted by fitting the $x$-behaviour
of  $\< h\vert T(J_{\rho L}(x) J^\dagger_{\rho L}(0))\vert h'\>$ to the
O.P.E. formula    
\begin{equation} \< h\vert T\Big (J_{\rho L}(x)
J^\dagger_{\rho L}(0)\Big)\vert h' \> = \sum_i c_i(x;\mu) \< h\vert
\widehat{O}^{(i)}(\mu)\vert h' \>  \,,
\label{eq:ME} 
\end{equation}
where the Wilson coefficients $c_i(x;\mu)$ are determined by continuum
perturbation theory using any renormalization scheme we like. The scale
$\mu$ should be chosen so that also the inequalities  $a\ll 1/\mu \ll 
\lqcd^{-1}$ are obeyed. Since we only consider operators of dimension 6 or
lower, the lattice $T$-product differs from the right-hand side of
eq.~(\ref{eq:ME}) by terms of $O(|x|^2\lqcd^2)$, which is then an estimate
of the size of the systematic errors intrinsic in this procedure. As a last
step we insert the fitted numbers $\< h\vert \widehat {O}^{(i)}(\mu)\vert
h'\>$ in~(\ref{eq:HEFFOPE}), obtaining in this way an explicit expression for
the matrix elements of  ${\cal H}_{\rm eff}^{W}$.

The procedure illustrated above requires the existence of a window, $a\ll |x| 
\ll  \lqcd^{-1}$, in which the distance between the two currents is
sufficiently small that perturbation theory can be trusted, but large enough
that lattice artifacts, which are suppressed by powers of $a/|x|$, are tiny.
For such a window to exist we need to have an adequately small lattice
spacing. At the same time the physical volume of the lattice must be
sufficiently large to allow the formation of hadrons.

A few remarks may be useful at this point.

$\bullet$ The method determines directly the ``physical" matrix elements of the
operators appearing in the O.P.E. of the  two currents, i.e. the matrix
elements of the finite, renormalized operators $\widehat{O}^{(i)}(\mu)$,
without any reference to the magnitude of the $W$-mass. This means that it
will not be necessary to probe distances of $O(1/M_{_W})$ with lattice
calculations. 

$\bullet$ Since it is the continuum O.P.E. which determines the operators appearing in 
the lattice expansion~(\ref{eq:HEFFOPE}), these are restricted by the
continuum symmetries. For $\lqcd^{-1}\gg |x|\gg a$, the lattice
O.P.E. matches, in fact, that of the continuum with discretization errors
suppressed by powers of $a/|x|$. 

$\bullet$ Unlike the methods discussed before, this approach automatically yields
hadronic amplitudes that are properly  normalized (in the renormalization
scheme in which the  Wilson coefficients appearing in eq.~(\ref{eq:ME}) are
computed).

As for the applicability of this strategy to the actual case of the
CP-conserving $\Delta S=1$ processes, fortunately in the
expansion~(\ref{eq:HEFFOPE}) there appear no operators of dimension lower
than 6. If lower  dimension operators were present they would dominate at
short distances, since their Wilson coefficients would diverge as powers of
$1/x$ (up to logarithmic corrections). In this situation it would be
virtually impossible to pick out the matrix elements of the interesting
dimension 6 operators.

Operators of dimension 6 have Wilson coefficients which vary
logarithmically with $|x|$. At leading order their expression is of the form
\be
c_i(x;\mu) \propto 
\left(\alpha_s(1/|x|) \over \alpha_s(\mu)\right)^{\gamma_0^{(i)}\over 2 
\beta_0} = 1 + \frac{\alpha_s}{4\pi}\gamma_0^{(i)} \log(|x| \mu) + \dots
\,,
\label{eq:formofci}
\ee
where $\gamma_0^{(i)}$ is the one-loop anomalous dimension of the operator
$O^{(i)}$ and $\beta_0$ is the coefficient of the one-loop term in the
$\beta$-function.

In the case of the $\Delta S=1$ part of ${\cal H}_{\rm eff}^{W}$, the
operators which can appear in~(\ref{eq:HEFFOPE}) are $\oplmi$ 
(eq.~(\ref{eq:oplmidef})) and in addition 
\be O' = (m_c^2-m_u^2) \,
\bar s (\overrightarrow{D_\mu} - \overleftarrow{D_\mu}) \gamma_\mu^L d
\,. \label{eq:prime}
\ee
The mass factor in the r.h.s. of eq.~(\ref{eq:prime}) comes from a
combination of the GIM mechanism, which causes $O'$ to vanish when
$m_c=m_u$, and chiral symmetry, which requires both quarks to be
left-handed, leading to a GIM factor  quadratic in the quark masses. Since
$O'$ has dimension 6, its coefficient function depends only logarithmically
on $|x|$.\par

The anomalous dimensions of the three relevant operators ($\oplmi$ and $O'$) 
are
\be
\gamma^{(+)}_0 = 4, \qquad 
\gamma^{(-)}_0 = -8, \qquad
\gamma '_0 = 16\ .
\label{eq:ANOMALDIM}
\ee

Actually, the contribution of $O'$ to the r.h.s. of eq.~(\ref{eq:ME})
can be determined separately (since its matrix elements do  
not require any subtraction) and needs not be fitted. The anomalous
dimensions of the operators $O^{(\pm)}$ are well separated from one another,
so it might be possible to determine the amplitudes $\< h\vert \oplmi
(\mu)\vert h' \>$ and obtain the physical matrix elements of ${\cal H}_{\rm
eff}^{\Delta S=1}$.

\section{CP-violating $\Delta S=1$ processes}
\label{sec:eps}

For CP-violating processes in kaon decays (or for $B$ decays), where
top-penguin diagrams enter at a Cabibbo-allowed level, the strategies
described at the points 1. and 2. of sec.~\ref{sec:deltai2} for the negative
parity operators fail, because the GIM mechanism is not operative, as the
top quark is too heavy to let it propagate on the lattice. In particular
$O^{(\pm)}$ mix with penguin operators (see below). This makes the
calculation of the mixing matrix of comparable difficulty to that for the
positive parity operators discussed at point 3. of sec.~\ref{sec:deltai2}.
For positive parity operators, the analysis carried out in
sec.~\ref{sec:deltai1} applies with the difference  that the mixing
coefficients of the color magnetic operator and scalar density become more
divergent~\cite{MMRT} and this will make the numerical determination
of the renormalized operators less precise. In order to circumvent these
problems we propose two methods involving a fictitious top quark (with mass
$\tilde m_t$) which is momentarily taken to be light enough to propagate on
the lattice.

\subsection{The Renormalization Group method}
\label{sec:rgope}

The basic idea of what we may call the ``Renormalization Group method"
is to work with two different scales: the first, $\mu$, is larger than
$\tilde m_t$, so that the corresponding operator basis is as in the previous
sections; the second, $\mu^\prime$, is smaller than $\tilde m_t$
so that a full set of new (penguin) operators is generated.  The matrix
elements of the operators renormalized at the scale $\mu$ are computed
numerically following  one of the strategies explained in
sec.~\ref{sec:deltai}. By matching the result to the amplitude expressed in
terms of operators renormalized at $\mu'$, we extract their matrix elements.
In this way, at least in principle, we can obtain the matrix elements of the
penguin operators without directly computing them. Let me now present
the details of this procedure.

At scales $a^{-1}\gg M_{_W}\gg\mu\gg\tilde m_t$, where the fake top quark is
active, the CP-violating part of the $\Delta S=1$ part of the effective weak
hamiltonian, ${\cal H}_{{\rm eff}}^{\Delta S=1}\vert_{\CPV}$, takes the form
(see eq.~(\ref{eq:HW})) 
\bea
 &\,&{\cal H}_{{\rm eff}}^{\Delta S=1}\vert_{\CPV}
= \lambda_t {G_F \over \sqrt2} 
[ C_1({\mu}, {M_{_W}}) (\widehat{O}^t_1(\mu)-\widehat{O}^c_1(\mu)) +
 \nonumber \\ &\,& 
+ C_2({\mu}, {M_{_W}}) (\widehat{O}^t_2(\mu)-\widehat{O}^c_2(\mu))] \,, 
\label{eq:HCPV}
\eea
\bea
 &\,&{O}^q_1 = (\bar s \gamma_\mu^L d)(\bar q \gamma_\mu^L q) \quad q=u,c,t
\\ &\,& {O}^q_2 = (\bar s \gamma_\mu^L q)(\bar q \gamma_\mu^L d) \quad q=u,c,t
\label{eq:HCPVOP}
\eea
where $\lambda_t=V_{td}V_{ts}^*$. The operators $\widehat{O}^{(\pm)}$ of
eq~(\ref{eq:oplmidef}) and the corresponding Wilson coefficients, $C_\pm$,
in eq~(\ref{eq:HW}) can be immediately written as linear combinations  of the 
operators and coefficients above.

At scales $\mu'$ below $\tilde m_t$, when GIM is not operative, the form of
the effective weak Hamiltonian becomes
\bea
{\cal H}_{{\rm eff}}^{\Delta S=1}\vert_{\CPV}&=&\frac {G_F} {\sqrt{2}}
\sum_{j=1}^N C_j(\mu',M_{_W},\tilde m_t) \widehat{O}_j(\mu')
\protect\label{eq:ehmu}
\eea
where now many new operators appear with a complicated mixing pattern with
lower dimensional operators~\footnote{A convenient basis of  operators when
QCD corrections are taken into account can be found in the papers quoted in
ref.~\cite{russi}. See also refs.~\cite{rev1}~\cite{rev2}.}. In fact in the
absence of GIM cancellation, in order to write a renormalized lattice
version of each of the operators $\widehat{O}_j$ it is required to subtract
lower dimensional operators with appropriate (power divergent) coefficients,
and account for the mixing with 6 dimensional operators. This is true for
both positive and negative parity sectors. The coefficient functions
appearing in eq.~(\ref{eq:ehmu}) have been  calculated up to non-leading
order in perturbation theory in refs.~\cite{noi}--\cite{ciuz}.

In order to avoid the difficulties mentioned above we have to keep the GIM
mechanism operative and this is done by having on the lattice a fictitious
top quark with mass satisfying 
\begin{equation}
1/a \gg \tilde{m}_t \gg m_c \gg \Lambda_{\rm QCD} \,.
\end{equation}

To illustrate how the method works in practice let me restrict the discussion
below to the simpler case of the negative parity  operators.
One begins by evaluating on the lattice at a renormalization scale
satisfying $\mu \gg \tilde{m}_t$ the two matrix elements 
\be 
{\cal M}_i(\mu,\tilde{m}_t) \equiv \langle h \vert 
 \widehat{O}^c_i(\mu) - \widehat{O}^t_i(\mu) 
\vert h^\prime \rangle \, ,\quad i=1,2 
\label{eq:match1}\ee
Since GIM is operative, the analysis of point 2. in 
sec.~\ref{sec:deltai2}
applies (with $m_u\to m_t$) and we can define  $\widehat{O}^c_i(\mu) -
\widehat{O}^t_i(\mu)$ in terms of bare lattice operators (by appropriately
subtracting the operator  $\bar s \gamma_5 d$, as discussed in
ref.~\cite{noinew}).

On the other hand, we can also consider the same matrix elements for a
renormalization scale $\tilde{m}_t\gg \mu' \gg m_c$. Since in this case the
GIM mechanism is not at work, all the operators which appear in 
eq.~(\ref{eq:ehmu}) can contribute and the amplitude ${\cal
M}_i(\mu,\tilde{m}_t)$ will take the form 
\be 
 {\cal M}_i(\mu,\tilde{m}_t) =
 \sum_{j=1}^N  \hat{Z}_{ij}(\mu^\prime,\mu, \tilde{m}_t) 
\langle h \vert \widehat{O}_j(\mu^\prime) \vert h^\prime \rangle \, .  
\label{eq:match2} 
\ee
The rectangular matrix $\hat{Z}_{ij}$ can be calculated
perturbatively by matching the theory with ($\mu\gg\tilde{m}_t$) and
without ($\mu '\ll\tilde{m}_t$) the fictitious top quark~\footnote{This is
exactly the method used to calculate the Wilson coefficients in the
theory with the physical top quark mass, except that in the physical case
one must simultaneously integrate out both $W$ boson and top quark.
Here we are effectively integrating  out the $W$ first and the fictitious top
quark afterwards.}. The results for $\hat Z$ at non-leading order can be 
reconstructed from the works of refs.~\cite{noi}--\cite{ciuz}. Notice that
in absence of QCD corrections one would have $C_1=0$, $C_2=1$.

We now choose a fixed value of $\mu^\prime$ and let $\tilde{m}_t$ vary.  The 
matrix elements of interest, $\langle h \vert \widehat{O}_j(\mu^\prime)
\vert h^\prime \rangle$, are obtained by fitting the $\tilde{m}_t$
dependence of the r.h.s. of eq.~(\ref{eq:match2}) to ${\cal
M}_i(\mu,\tilde{m}_t)$, computed numerically as in eq.~(\ref{eq:match1}), and
using the renormalization matrix $\hat{Z}_{ij}$ calculated perturbatively.
Since the dependence on $\tilde{m}_t$ is only logarithmic, this will not be
an easy job. The procedure is analogous to our use of the $x$-dependence in
sec.~\ref{sec:ope} to find the renormalized matrix elements of operators
appearing in the weak Hamiltonian. Having determined the numbers
$\langle h \vert \widehat{O}_j(\mu^\prime) \vert h^\prime \rangle$, we can
insert them into the expression (\ref{eq:ehmu}) for the effective
Hamiltonian.  At this point the constraint $M_{_W} \gg \tilde{m}_t$ can be
removed since the Wilson coefficients of the operators appearing
in~(\ref{eq:HCPV}) can be computed perturbatively for arbitrary values of
$\tilde{m}_t$, including $\tilde{m}_t=m_t$.

As for the errors involved in this procedure, it should be observed that for
an accurate determination of the matrix elements of $\widehat{O}_i(\mu)$ (or
of $\widehat{O}_j(\mu^\prime)$) the condition
$\mu\gg\tilde{m}_t\gg\mu'\gg\Lambda_{\rm QCD}$ is not enough. It must also
be required that the typical scale, $\Lambda_{hh'}$, of masses and external
momenta appearing in the physical process $h^\prime \to h$ be much smaller
than $\tilde m_t$. This is because in the matching procedure terms of
$O(\Lambda_{hh'}/\tilde m_t)$ are neglected. 

\subsection{The O.P.E. method}
\label{sec:opemethod}

An alternative method, in the same spirit of the approach followed in
sec.~\ref{sec:ope}, is the following.  We can avoid the need for any
non-perturbative subtraction by separating the two currents in $x$ space and
having a fictitious propagating top quark. Thus, as in sec.~\ref{sec:ope}, we
directly match the $x$-behaviour of $\langle h\vert T(J_{\rho L}(x)
J^\dagger_{\rho L}(0)) \vert h'\rangle_{\rm top}^{\Delta S = 1}$ (the 
subscript indicates the presence of the fictitious top) to the formula
\begin{equation}
\langle  h\vert T(J_{\rho L}(x) J^\dagger_{\rho L}(0))
\vert h' \rangle_{\rm top}^{\Delta S = 1} = \sum_{j}
c_j(x; \mu^\prime, \tilde{m}_t) \langle h\vert
 \widehat{O}^{(j)}(\mu^\prime)\vert h' \rangle
\,,\label{eq:MEtop}
\end{equation}
where 
\be J_{\rho L}(x)  J^\dagger_{\rho L}(0)\vert^{\Delta S = 1}= 
 \bar s(x) \gamma_{\mu}^L t(x)
\bar t(0) \gamma_{\mu}^L d(0) -
 \bar s(x) \gamma_{\mu}^L c(x)
\bar c(0) \gamma_{\mu}^L d(0) \, .\ee
The coefficients $c_{1,2}(x;\mu^\prime, \tilde{m}_t)\equiv
c_{1,2}(x;\mu^\prime)$ are the same as those of sec.~\ref{sec:deltai}. The
others are complicated functions of the anomalous dimension matrix which can 
be worked out from the results of refs.~\cite{buras2} and~\cite{noi2} and 
computed numerically.

Both methods proposed in this section require small enough lattice
spacings to accommodate a number of scales. Like the method of
sec.~\ref{sec:ope} their full implementation is likely to require the
next generation of supercomputers.

\section{Conclusions}
\label{sec:concs}

The new experimental results on $\epsilon'/\epsilon$~\cite{epsepsn}, combined
with the old ones~\cite{epsepso}, have prompted a wage of theoretical work
aimed at testing the validity of the Standard Model in this crucial corner
of the theory. Lattice, as shown recently by the investigation carried out
in ref.~\cite{col}, has the potentiality of providing a clearcut answer to
the question on whether CP-violation can be understood and described within
the framework of the Standard Model, as it is formulated today. In my
opinion all possible efforts should be addressed by the lattice community to
this fundamental issue.

With this in mind in this talk I have reviewed a number of old and new
approaches aimed at studying  the \dirule\ on the lattice, using Wilson-like
fermions (similar methods could also be used for staggered
fermions~\cite{toolkit}) and I have discussed a new approach which can be
equally well applied to CP-conserving and CP-violating $\Delta S =1$
processes. 

In particular the methods of refs.~\cite{direct} and~\cite{MMRT} involving
$K\to\pi\pi$ and $K\to\pi$ amplitudes, respectively have been reevaluated.
The last approach is likely to be more difficult than the first one, because
of the large number of mixing coefficients which have to be determined
non-perturbatively. It may however give complementary information to the
results obtained with the $K\to\pi\pi$ methods (points 1. and 2. of
sec.~\ref{sec:deltai2}) and provide a check of the accuracy of chiral 
extrapolations.

To overcome the difficulties inherent in the construction of finite
renormalized lattice operators, a new strategy, based on the
study of the short distance behaviour of the O.P.E. of two hadronic weak
currents on the lattice, was proposed in ref.~\cite{noinew}. The approach is
theoretically very appealing and can be applied to both CP-conserving and
CP-violating $\Delta S =1$ processes. There are also good
indications~\cite{sch} that the approach can be extended in a way as to
allow a fully non-perturbative evaluation of the (first few) moments of
D.I.S. structure functions~\footnote{The generalization proposed in
ref.~\cite{sch} makes use of (gauge non-invariant) quark and gluon
states in order to get a sufficiently large set of equations and be able to
extract all the unknown coefficients and matrix elements. In these
circumstances non-BRST operators vanishing by the equations of motion can
contribute to the O.P.E.~\cite{NBRST}. If they are not included, care must be
exerced in choosing the external states in order to avoid contaminations from
these unwanted operators onto the relevant matrix elements.}.

An other interesting feasibility study in the direction of testing the O.P.E.
method in actual simulations has been undertaken in ref.~\cite{MONT}. It
consists in the study of the small $x$-behaviour of the O.P.E. in the
two-dimensional lattice O(3) $\sigma$-model. A preliminary analysis of Monte
Carlo data indicates that the measured small $x$-behaviour of the
one-particle matrix elements of products of operators matches the
logarithmic behaviour expected from  perturbative calculations, thus
allowing the (non-perturbative) evaluation of the matrix elements of the
operators appearing in the O.P.E.

The advent of Teraflop Supercomputers may render the  O.P.E. method a viable
strategy also for QCD, allowing us to directly extract physical amplitudes
from Monte Carlo data, which may compete with the more standard approaches
illustrated in sec.~\ref{sec:deltai} or with those, based on Ginsparg-Wilson,
overlap or domain wall fermions~\cite{GWDWF}, that are now starting
to be developed.

\section*{Acknowledgements}

I would like to thank the Organizers of ``CHIRAL '99", and especially Prof.
Ting-Wai Chiu, for the warm hospitality and for the very exciting
scientific atmosphere that they have been able to create during the meeting.

\end{document}